\documentclass[aps,amssymb,amsmath,pra, 12pt,showpacs, superscriptaddress]{revtex4}
\usepackage{amsfonts}
\usepackage{amsmath}
\pdfoutput=1
\usepackage{graphicx}
\usepackage{epsfig}

\begin{document}

\title{A New Hierarchical Genetic Algorithm Approach to Determine Pulse Sequences in NMR}
\author{Ashok Ajoy}
\email{ashok.ajoy@gmail.com}
\affiliation{Birla Institute of Technology and Science - Pilani, Zuarinagar, Goa - 403726, India.}
\affiliation{NMR Research Centre,
Indian Institute of Science, Bangalore - 560012, India.}
\author{Anil Kumar}
\affiliation{NMR Research Centre,
Indian Institute of Science, Bangalore - 560012, India.}

\begin{abstract}
Nuclear Magnetic Resonance (NMR) spectroscopy provides a valuable tool by which one can control a spin ensemble. Control is achieved by using radio-frequency (RF) pulses. Pulse sequence design has been an active research area for many years. Recently, optimal control theory has been successfully applied to the design of pulse sequences, so as to minimize their total duration and improve their efficiency.\\
In this paper, we develop a new class of genetic algorithm that computationally determines efficient pulse sequences to implement a quantum gate $U$ in a three-qubit system. The method is shown to be quite general, and the same algorithm can be used to derive efficient sequences for a variety of target matrices. We demonstrate this by implementing the inversion-on-equality gate efficiently when the spin-spin coupling constants $J_{12}=J_{23}=J$ and $J_{13}=0$. We also propose new pulse sequences to implement the parity gate and fanout gate, which are about 50\% more efficient than the previous best efforts. Moreover, these sequences are shown to require significantly less RF power for their implementation.\\
The proposed algorithm introduces several new features in the conventional genetic algorithm framework. We use matrices instead of linear chains, and the columns of these matrices have a well defined hierarchy. The algorithm is a genetic algorithm coupled to a fast local optimizer, and is hence a hybrid GA. It shows fast convergence, and running on a MATLAB platform takes about 20 minutes on a standard personal computer to derive efficient pulse sequences for any target 8X8 matrix $U$.
\end{abstract}
\maketitle

\section{Introduction}
In recent years, there has been considerable interest in formulating time optimal pulse sequences in NMR. Various efforts have focused on replacing traditionally well known sequences (for example, sequences to transfer coherence between coupled spins in multidimensional NMR experiments \cite{kha01}) by their time optimal counterparts.\\
The advantages of time optimal sequences are many. By reducing the time required to perform a desired unitary operation, they reduce the impact of undesirable effects due to decoherence or relaxation. The efficiency in achieving the desired operation can be improved drastically (in some cases it can be doubled \cite{kha02}). It is becoming clear that any serious attempts at quantum computing \cite{qc1} using NMR would require such time optimal sequences at their foundation.\\
The process of formulating time optimal sequences, like every other process of optimization, involves minimizing a "cost" function. The most widely used cost function for a coupled spin system is the time for evolution of the system under spin-spin J coupling [1,2]. Under the spin diffusion limit approximation, the time required to implement hard pulses is negligible compared to this time.\\
The NMR Hamiltonian can be decomposed as [1]
\begin{equation}
H = H_d + \sum_{j=1}^{m} v_jH_j
\end{equation}
where drift term $H_d$ is the part of the Hamiltonian internal to the system, consisting of the spin-spin coupling term. $\sum_{j=1}^{m} v_jH_j$, is the part of the Hamiltonian that can be externally changed (this is achieved by using hard pulses). In this paper, we shall deal with the three-spin problem, where the spin-spin coupling constants $J_{12}=J_{23}=J$, and $J_{13}=0$ [2]. In this case,
\begin{eqnarray*}
H_d &=& 2\pi J(I_{1z}I_{2z} + I_{2z}I_{3z})\\
H_1 &=& 2\pi I_{1x}\\
H_2 &=& 2\pi I_{1y}\\
H_3 &=& 2\pi I_{2x}\\
H_4 &=& 2\pi I_{2y}\\
H_5 &=& 2\pi I_{3x}\\
H_6 &=& 2\pi I_{3y}\\
\end{eqnarray*}
The unitary evolution of 3 interacting spin $\frac{1}{2}$ particles is described by an element of SU(8), the group of unitary matrices \cite{hel} with determinant 1. The Lie algebra su(8) is a 63 dimensional space whose basis operators are 8x8 skew-Hermitian matrices [2,4]. Clearly then, the NMR time-optimal problem in this case becomes an optimization problem in a 63 dimensional space.\\
Recent efforts have used geometric control theory, where the time optimal sequence is derived from the optimal trajectory (geodesic)\cite{kha02,kha09} to be traversed from the initial state to the desired final state in this 63 dimensional space. For example, trilinear propagators of the form $U=\exp(-i\theta I_{1\alpha}I_{2\beta}I_{3\gamma})$, where $\alpha, \beta, \gamma \in \{x,y,z\}$, have been implemented using geodesic sequences having total period of about half of traditional methods.\\
There have also been pulses optimized using dynamic programming \cite{relax1}. This is a so called "greedy hill climbing" optimization technique. Essentially, any optimization problem can be viewed as consisting of a fitness landscape, where the goal of optimization is to reach to highest hill(global optimization). Dynamic programming is a point to point technique where one traverses from one point of the landscape to another point having strictly higher fitness. This has been used to optimize coherence transfer in the presence of relaxation (the ROPE sequence) \cite{relax1,relax2,relax3}.\\
However, finding time-optimal sequences for a general 3 spin NMR problem has still remained unsolved. Here it is desired to form a desired 8x8 unitary operator using pulse sequences that require minimum period of evolution under the drift Hamiltionian $H_d$ (spin-spin coupling). We seek to solve this problem in two steps: first given a unitary operator $U$, we find many (if not all) possible ways of realizing this operator using state-of-art optimal sequences \cite{kha02}. Then we find which of these sequences is the best. By automating this process and devising an algorithm to decompose the operator $U$ using optimal sub-sequences, it is hoped that we can obtain more efficient sequences for $\it{any}$ 8x8 operator $U$.\\

The main results of this paper are as follows:
\begin{enumerate}
\item We formulate a new genetic algorithm that provides various possible sequences for a matrix $U$ using cascaded time optimal sequences. This genetic algorithm is different from conventional implementations, and has added features of being a hybrid with a local-optimizer, and having an in-built hierarchy that makes it faster.\\
\item We then use this algorithm to optimize the parity gate and fanout gate. The new sequences are found to 50\% more efficient than exisitng sequences, besides requiring much less RF power for their implementation \cite{anil1,anil2}. In case of the $U_{=}$ invert on equality gate, we determine a new effienct sequence when the spin-spin coupling $J_{13}=0$ and $J_{12}=J_{23}$.\\
\end{enumerate}

\section{Product Operator basis in NMR}
We will only be interested in the 3-spin NMR problem that forms the space $SU(8)$. Consider the Pauli matrices defined by \cite{ernst}:

\begin{equation}
I_x = \frac{1}{2}\left(\begin{array}{cc} 0 & 1 \\ 1  & 0 \end{array}\right), I_y =\frac{1}{2}\left(\begin{array}{cc} 0 & -i \\ i  & 0 \end{array}\right), I_z = \frac{1}{2}\left(\begin{array}{cc} 1 & 0 \\ 0  & -1 \end{array}\right)\\
\end{equation}
They obey the well known commutation relations
\begin{eqnarray}
[I_x,I_y] = iI_z; [I_y,I_z] = iI_x; [I_z,I_x] = iI_y \\
I_x^2 = I_y^2 =I_z^2 = \frac{1}{4}\bf{1}
\end{eqnarray}
where $\bf{1}$ is the identity element. The product operator basis is an orthogonal basis $iB_s$, which for an n-spin problem [$su(2^n)$] takes the form
\[B_s = 2^{q-1}\prod_{k=1}^{n} (I_{k\alpha})^{a_{ks}}\]
where $\alpha=\{x,y,z\}$ and $I_{k\alpha} = 1\otimes\cdots\otimes I_{\alpha}$.\\
Hence for a 3-spin problem, there are 64 base operators including the unity operator $\bf{1}$. These can be conveniently put in the tabular form as in Fig 1

\begin{figure}
\includegraphics[width=16cm]{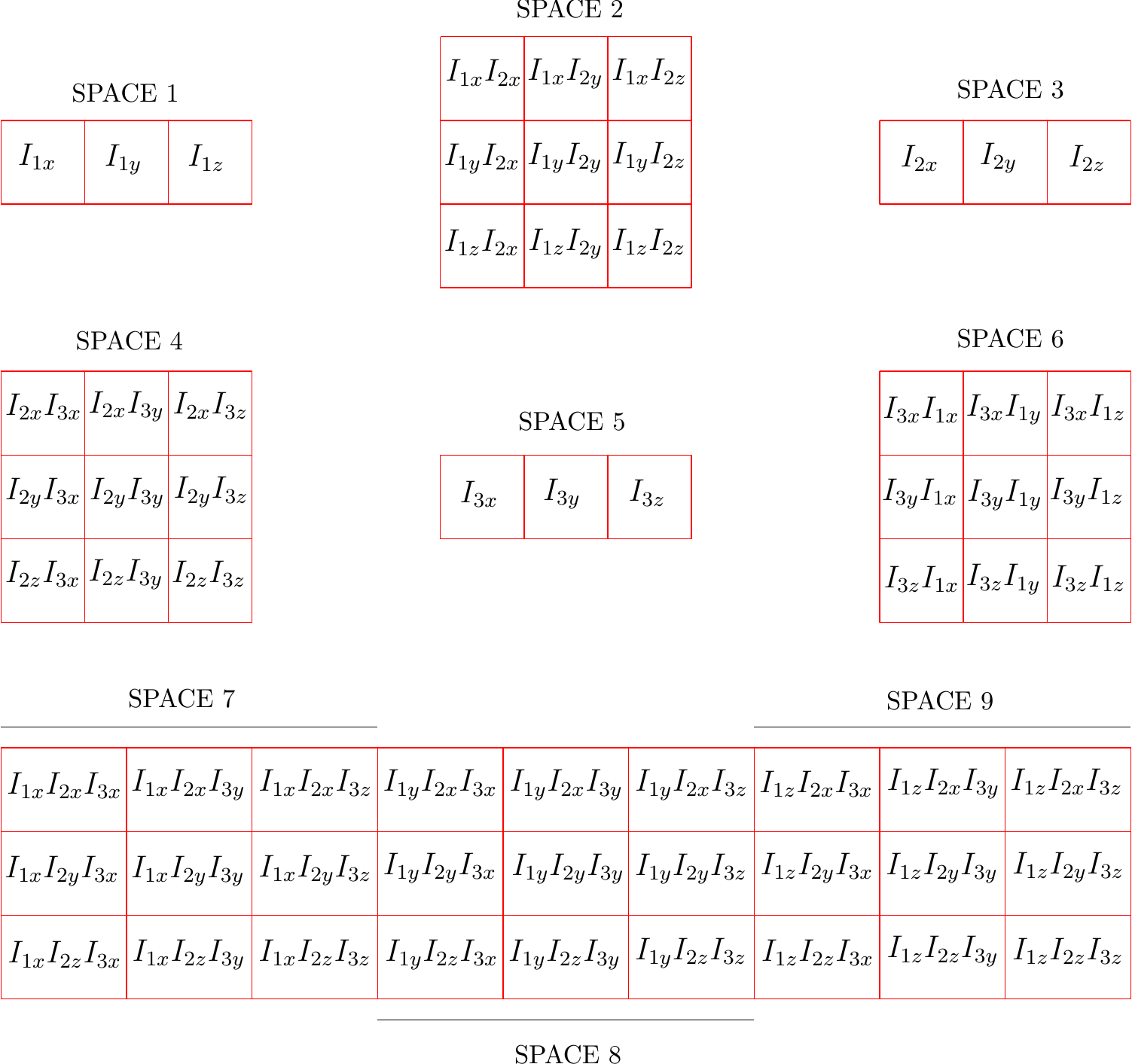}
\caption{The panel illustrates the 63 base operators of the product operator basis. Along with the unity operator $\bf{1}$, they form a basis to describe any 8x8 matrix for a 3-spin problem. The separated boxes show the different subspaces, and it is possible to travel inside each subspace by using only hard pulses, which in our optimization problem has no cost.}
\end{figure}

This table is a way of representing the 63 base operators that form a basis (along with the identity operator $\bf{1}$) in su(8). The Lie algebra su(8) can be decomposed as 
\begin {equation}
\rm{su(8)} = \mathfrak{l_1} \oplus \mathfrak {l_2} \oplus \mathfrak {l_3} \oplus \mathfrak {p_1} \oplus \mathfrak {p_2} \oplus \mathfrak {p_3} \oplus \mathfrak {q}
\end{equation}
where
\begin{eqnarray*}
\mathfrak{l_1} = &=& \rm{span}\ i\{I_{1x},I_{1y}, I_{1z}\}\\
\mathfrak{l_2} = &=& \rm{span}\ i\{I_{2x},I_{2y}, I_{2z}\}\\
\mathfrak{l_3} = &=& \rm{span}\ i\{I_{3x},I_{3y}, I_{3z}\}
\end{eqnarray*}
represent the subspaces that are spanned by the hard-pulses on the first, second and third spins respectively. Bilinear spin terms form the subspace $\mathfrak {p_1} \oplus \mathfrak {p_2} \oplus \mathfrak {p_3}$, where
\begin{eqnarray*}
\mathfrak{p_1}  &=& \rm{span}\ i\{2I_{1x}I_{2x},2I_{1x}I_{2y},2I_{1x}I_{2z},2I_{1y}I_{2x},2I_{1y}I_{2y},2I_{1y}I_{2z},2I_{1z}I_{2x},2I_{1z}I_{2y},2I_{1z}I_{2z},\}\\
\mathfrak{p_2}  &=& \rm{span}\ i\{2I_{2x}I_{3x},\cdots\}\\
\mathfrak{p_3}  &=& \rm{span}\ i\{2I_{3x}I_{1x}, \cdots\}
\end{eqnarray*}
where the $\cdots$ in the last two equations represent 8 other operators ( a total of 9) formed by cyclic permutations of $\{x,y,z\}$. \\
The subspace $\mathfrak{q}$ contains the trilinear terms
\begin{equation}
\mathfrak{q}= \rm{span}\ i\{4I_{1x}I_{2x}I_{3x}, 4I_{1x}I_{2x}I_{3y}, 4I_{1x}I_{2x}I_{3z}, \cdots\}
\end{equation}
where the $\cdots$ represents a total of 27 operators whose span forms $\mathfrak{q}$, and which are formed by the cyclic permutation of the indices $\{x,y,z\}$ in the trilinear propagator.\\

It is this decomposition of $su(8)$ that table 1 represents. The key issue involved with such a decomposition is the immediate utility it has with regards to our optimization problem. The table shows that are a total of 7 subspaces that form su(8), but \textit{we can move within each subspace by using only hard pulses} \cite{kha01,kha02}. For example, any trilinear propagator $\exp(-i\theta I_{1\alpha}I_{2\beta}I_{3\gamma})$  where $\alpha, \beta, \gamma \in \{x,y,z\}$ can be implemented from $\exp(-i\theta I_{1z}I_{2z}I_{3z})$ by using only hard pulses. In our optimization problem , hard pulses have no "cost", and hence all 27 trilinear propagators are equivalent in their fitness. This reasoning is also true for the bilinear propagators, ie any propagator $\exp(-i\theta I_{n\alpha}I_{l\beta})$ where $\alpha, \beta\in \{x,y,z\}$ and $n,l \in \{1,2,3\}$, can be implemented from $\exp(-i\theta I_{nz}I_{lz})$ by using only hard pulses. Two examples of the above ideas are \cite{kha01,kha02}:
\begin{eqnarray}
\exp(-i\theta I_{1x}I_{2y}I_{3z}) &=& \exp\left(-i\frac{\pi}{2}I_{1y}\right)\exp\left(i\frac{\pi}{2}I_{2x}\right)\exp(-i\theta I_{1z}I_{2z}I_{3z})\exp\left(-i\frac{\pi}{2}I_{2x}\right)\exp\left(i\frac{\pi}{2}I_{1y}\right)\nonumber\\
\exp(-i\theta I_{1y}I_{2y}) &=& \exp\left(i\frac{\pi}{2}I_{1x}\right)\exp\left(i\frac{\pi}{2}I_{2x}\right)\exp(-i\theta I_{1y}I_{2y})\exp\left(-i\frac{\pi}{2}I_{2x}\right)\exp\left(-i\frac{\pi}{2}I_{1x}\right)
\end{eqnarray}

In summary any unitary 8x8 matrix $U$ can be decomposed using hard-pulses, bilinear propagators of form $\exp(-i\theta I_{n\alpha}I_{l\beta})$ and trilinear propagators of form $\exp(-i\theta I_{1\alpha}I_{2\beta}I_{3\gamma})$. As far as the cost during the optimization process is concerned, all hard pulses have zero cost \cite{kha01,kha02,kha07,kha09}, all bilinear propagators are equivalent and all trilinear propagators are equivalent (for the same phase angle $\theta$).

\section{The NMR Toolbox}
In literature, trilinear propagators have been implemented optimally using geodesic pulse sequences. Bilinear propagators are implemented by standard refocusing techniques, where one of the spins is decoupled from the system. These sequences are shown in fig 2.

\begin{figure}
\includegraphics[width=8cm]{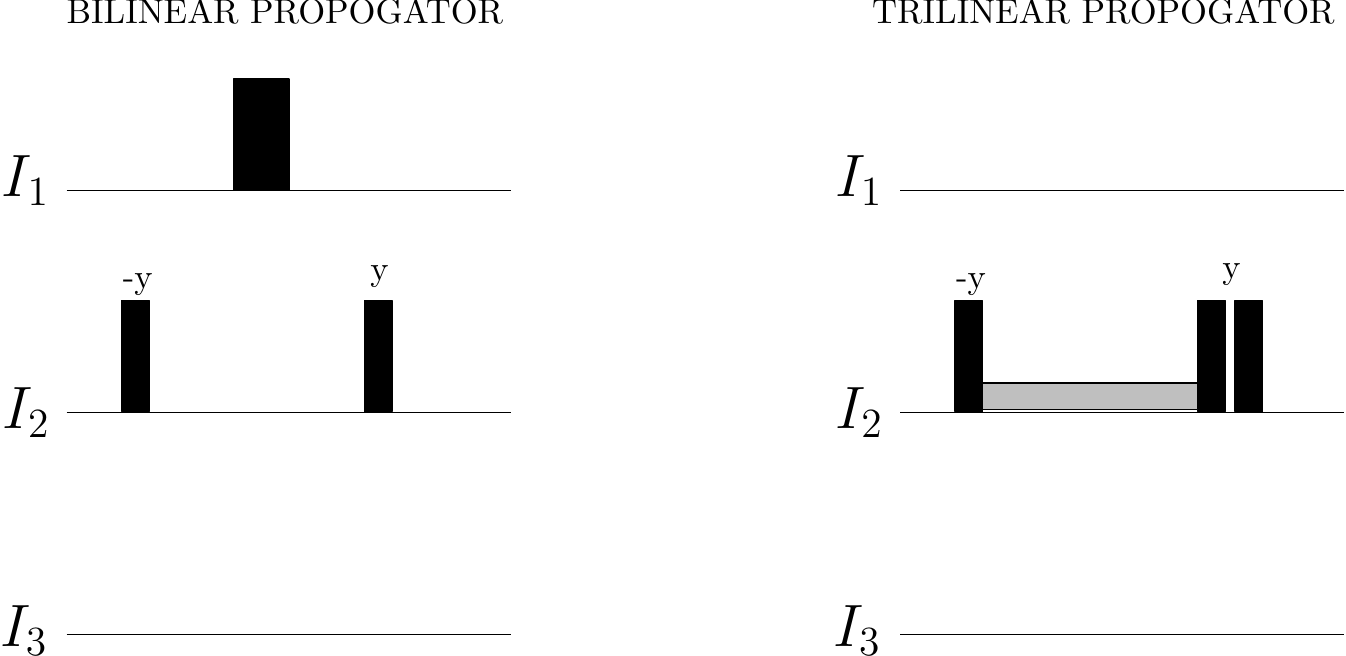}
\caption{The panel illustrates the refocusing method of implementing bilinear propagators, and the geodesic sequence for implementing trilinear propagators}
\end{figure}

Given a target unitary matrix $U$, we now break the problem of finding time-optimal pulses sequences to realize $U$ into 2 steps
\begin{enumerate}
\item Devise an algorithm to decompose $U$ in several ways, using only hard pulses, bilinear and trilinear propagators. For a 3 spin problem consisting of a 63-dimensional space, there may be several ways to realizing the matrix $U$. Consider a N-step decomposition of U as
\begin{equation}
U=\exp(-i\theta_1 I_{n\alpha})\exp(-i\theta_2 I_{l\beta}I_{m\gamma})\exp(-i\theta_3 I_{p\delta}I_{q\epsilon}I_{r\eta})\cdots (N steps)
\end{equation}
where the greek letters $\alpha,\beta\cdots \in \{x,y,z\}$, and roman letters $n,l\cdots \in \{1,2,3\}$, and any of $\theta_i, i\in [1,N]$ may be 0.\\
At the outset this problem seems complicated, because not only are the angles $\theta_i$ variable, but so also are the choice of propagator and their ordering (because succeeding propagators may not commute). However, we develop a hybrid genetic algorithm, that is surprisingly fast in converging to a solution. (sec 5). We sweep through the number of steps $N$ from about 3 to 10 to find many (if not all) possible ways of decomposing the matrix $U$. \\
\item We then simply choose the best sequence amongst these sequences, and this is likely to be the best way of achieving the target $U$.
\end{enumerate}

\section {Genetic Algorithms}
Genetic Algorithms (GAs) are a class of global optimization algorithms \cite{ga1,ga2,ga3} first introduced by John Holland in the 70s, and have be applied to various areas from mechanical engineering to radio-astronomy. They are biologically inspired algorithms, and seek to mimic the process of evolution and the strategy of "survival of the fittest". The key components of a traditional GA are a collection of bit-strings \cite{ga2} that form a "population". The bit strings are usually coded representations of the various candidate solutions to the problem at hand, and the coding scheme is flexible and differs from problem to problem. Usually the starting population is a random selection from the candidate space \cite{ga1}. Then, mimicking the process of mating and evolution, pairs of strings are chosen from the population and \textit{crossed over}. Crossing over entails exchanging the bit-strings after a certain bit-position called a locus \cite{ga2} (this is randomly chosen). However, similar to the "survival of the fittest" adage, the member of the population that is fittest is more likely to cross-over (a member can cross over more than once). There is also another operation called a \textit{mutation}, which involves flipping over a bit at a random location \cite{ga2,ga3}. Mutation and cross-over are competing operations, but the probability of mutation is usually kept low (about 10\% chance of mutation and 90\% chance of cross-over). If this process is continued over several reporduction cycles (generations), it is found that the population becomes fitter and fitter \cite{ga1}, until finally the optimum is reached. Fig 3 shows the tradional cross-over and mutation operations, and a flow-chart implementation of a simple GA.\\
\begin{figure}
\includegraphics[width=8cm]{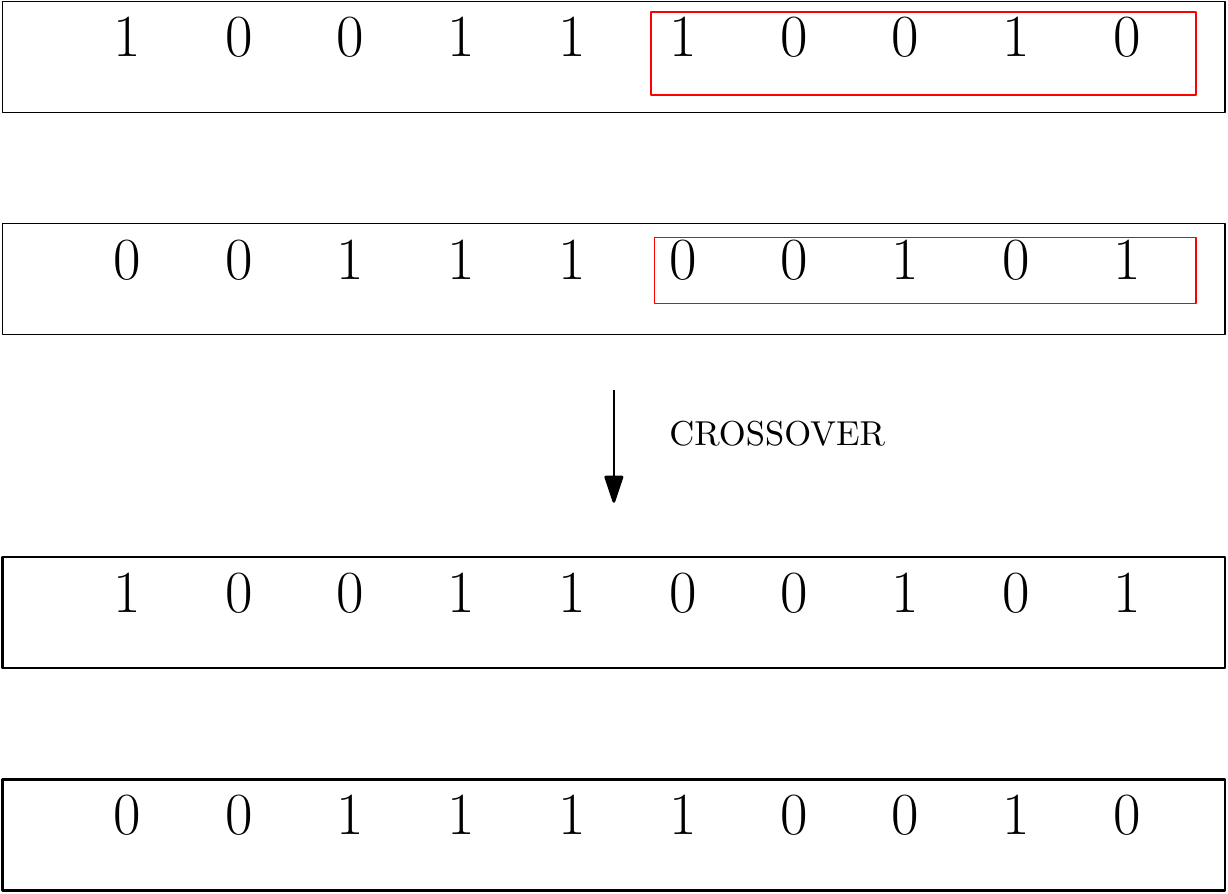}
\caption{Cross over operation in a traditional genetic algorithm. Here cross over is taking place at a locus =6}
\end{figure}

\begin{figure}
\includegraphics[width=8cm]{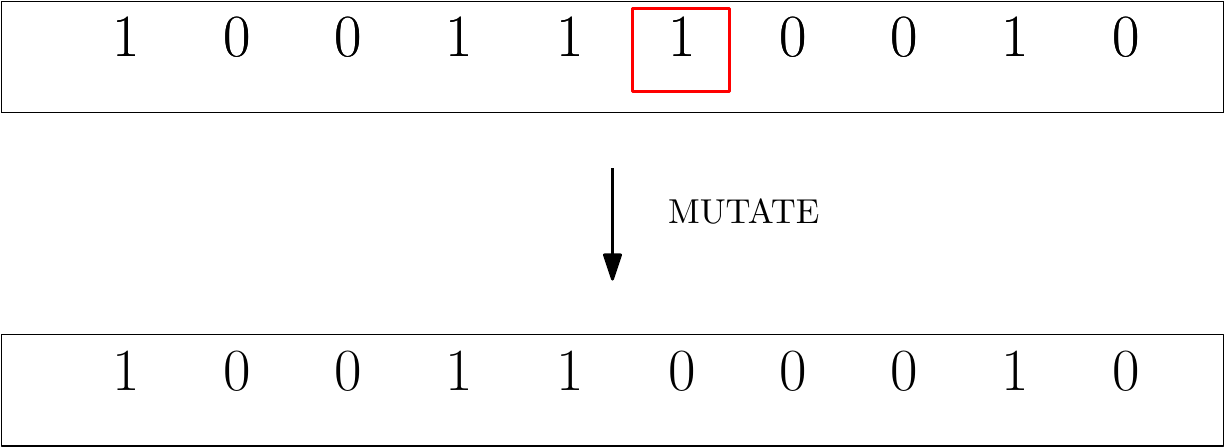}
\caption{Mutation operation in a traditional genetic algorithm. Here mutation is taking place at a locus =6}
\end{figure}

The main advantage of GAs are that they are a parallel search technique \cite{ga1}, where all the members of the population search the candidate space simultaneously. Moreover, unlike point-to-point optimization methods like dynamic programming, where we always seek points of higher fitness, genetic algorithms allocate in a principled way a limited number of trials to solutions that are known to be inferior. It is this diversity, that potentially leads to the GA approaching the global optimum where other methods may fail. However, on the flip side, due to their high reliance on randomness and diversity, GAs are usually slow to converge to a solution.\\

\section {New hybrid hierarchical class of genetic algorithm}
In our problem, we want to formulate a GA that provides several (if not all) possible decompositions of a target matrix $U$ into hard pulses, bilinear and trilinear propagators. Because of the large number of variables, and problems due to non-commutativity, convergence of the GA in a reasonable time-frame would become difficult. Hence we break away from the traditional framework of the GA, and introduce new features that respect the basis of genetic algorithms (parallel search based on fitter members crossing over more), but improve convergence:
\begin{enumerate}
\item We use matrices instead of bit strings\\
\item We use digits 0-9 instead of restricting to only 0 or 1.\\
\item Our encoding scheme is as follows. The matrix consists of 4 columns and a variable number of rows. The number of rows specify the number of steps ($N$) we want to decompose the matrix $U$ into. 
\begin{itemize}
\item The first column represents the subspace from which to pick a propagator (in table 1), note that this can go from 1-9.\\
\item The second column represents the propagator inside that subspace. This corresponds to the number of the square inside the subspace in fig 1. \\
For example if a row has as first two elements 4 and 7, then it would corresponding to the propagator in the 7th square in the 4th subspace in fig 1, and this is $\exp(-i\theta I_{2z}I_{3y})$. For the subspaces 1,3,5 in fig 1 where there are only 3 propagators, we make them triplicate, ie numbers 1-3, 4-6, and 7-9 in the second column would all be the same. This is to standardize the encoding scheme to have 9 digits.
\item The third and fourth columns represent the angle of rotation $\theta$ of the selected propagator. The angle is computed as:
\begin{equation}
\theta= \textrm{(3rd column number)}\times 45^{\circ} + \textrm{(4th column number)}\times 5^{\circ}
\end{equation}
\end{itemize}
\item Hence there is a certain \textit{hierarchy} built into this encoding scheme. A change in number in the 4th column is unlikely to affect the solution as much as a change in the number in the 3rd column.\\
\item This encoding scheme ensures that the candidate solution converges to within a 5$^{\circ}$ accuracy in each of the $\theta_i, i\in [1,N]$. We then use a local optimizer, (which can be a modified genetic algorithm, or a dynamic programming algorithm) to approach to within an arbitrary accuracy (upto $0.1^{\circ}$). It is thus we call the algorithm a \textit{hybrid} genetic algorithm, where we couple the inherent global optimization advantages of a GA with the fast convergence of a local-optimizer.
\end{enumerate}

\subsection{Examples of Encoding Scheme}
To make things clear, consider two 4x4 matrices in Fig that are members of a population at some time. From Fig 1 and using our encoding scheme they correspond to the matrix decomposition:
\begin{eqnarray}
A&=& \exp\left(-i\frac{95\pi}{180}I_{1z}\right)\exp\left(-i\frac{85\pi}{180}I_{1z}I_{2y}I_{3y}\right)\nonumber\\
&\times&\exp\left(-i\frac{285\pi}{180}I_{1x}I_{2z}\right)\exp\left(-i\frac{175\pi}{180}I_{1x}I_{2x}I_{3y}\right)\\
B&=&\exp\left(-i\frac{435\pi}{180}I_{1x}I_{2y}I_{3z}\right)\exp\left(-i\frac{280\pi}{180}I_{1x}I_{2x}I_{3y}\right)\nonumber\\
&\times&\exp\left(-i\frac{300\pi}{180}I_{1z}I_{2y}\right)\exp\left(-i\frac{190\pi}{180}I_{1z}I_{2x}I_{3z}\right)
\end{eqnarray}
\begin{figure}
\includegraphics[width=8cm]{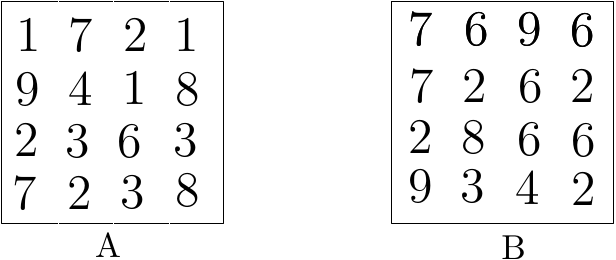}
\caption{Two examples of the encoding scheme. The sequences they correspond to are given below.}
\end{figure}

\subsection{Crossover and Mutation operations}
\begin{figure}
\includegraphics[width=12cm]{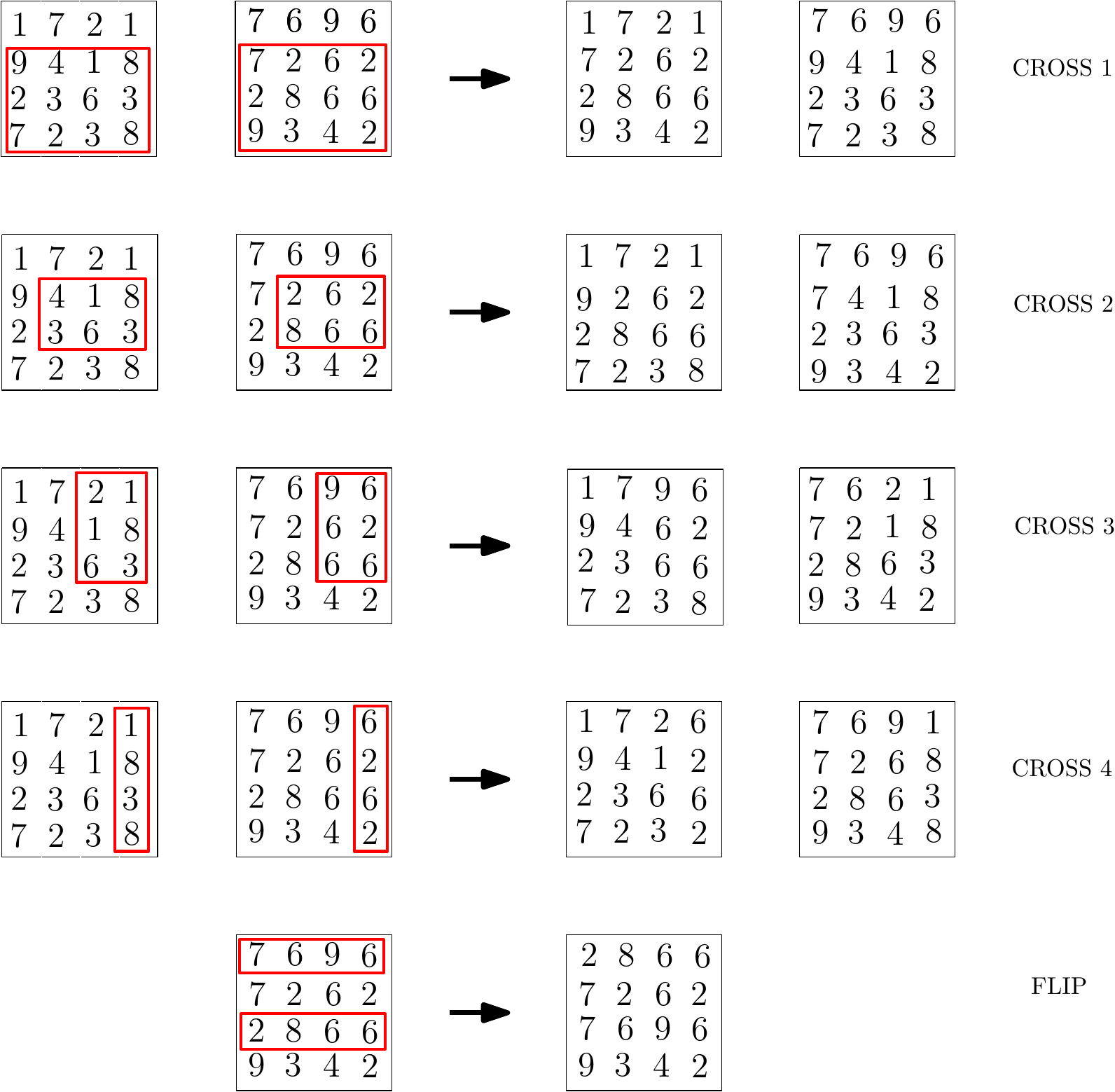}
\caption{Crossover operations in our genetic algorithm, applied to the two example population members A and B above. There are 4 kinds of cross-over operations and 1 flip operation to account for non-commutativity amongst succeeding propagators. The CROSS operations involve two members, while the FLIP operation requires only one member.}
\end{figure}
Traditionally, crossover involves exchanging 2 chosen bit strings at a randomly chosen locus (cross-over point), and repeating this process till a new population is created. In our scheme, crossover involves exchange of subblocks between 2 matrices (members of the population). Since we must respect the hierarchy between columns, we introduce 4 different kinds of crossover operations:
\begin{enumerate}
\item \textit{Cross 1} : Involves cross over of sub-blocks of size $n\times 4$ where $n\in [1,N]$, and $N$ is the total number of rows of each member.
\item \textit{Cross 2} : Involves cross over of sub-blocks of size $n\times 3$ where $n\in [1,N]$, and the first column is not involved in cross-over. This is a means of respecting hierarchy, because the first column is more important than the second, which in turn is more important than the third etc. Additionally, at the end of this operation, with a small probability (10\%), a \textit{mutation} can be introduced in the first column. In thise case, a random entry is chosen in the first column and is randomized in 1-9.
\item \textit{Cross 3} : Involves cross over of sub-blocks of size $n\times 2$ where $n\in [1,N]$, and the first two columns are not involved in cross-over. With a probability of 7\%, a \textit{mutation} can be introduced in the first two columns.
\item \textit{Cross 4} : Involves cross over of sub-blocks of size $n\times 1$ where $n\in [1,N]$, and the first three columns are not involved in cross-over. With a probability of 50\%, a \textit{mutation} can be introduced in the first three columns.
\item \textit{Flip} : This is an operation that seeks to address problems due to non-commutativity of successive propagators, and hence to improve convergence to a valid decomposition. This is a unary operation, and involves only one member of the population (unlike cross-over that occurs between two members). Here, two rows are randomly selected and flipped over (exchanged). This can be done more than once. This process is repeated to half the total members of the population .
\item \textit{Mutate} : Genetic algorithms traditionally have a tendency to get trapped at local optima. This is because a certain member of the population may be fitter than the remaining (but not the fittest possible), and may reproduce widely, leading to all members having almost the same characteristics, and the GA getting "trapped". To break out of this scenario, we introduce the \textit{mutation} operation after the 30th generation. The probability of mutation increases linearly from 0 in the 30th generation to about 0.35 in the 50th generation (we run upto 50 generations). In this operation, 10 members of the population (usually population size is taken between 500 and 1000) are chosen, and all their entires are randomized between 0-9.
\end{enumerate}

\subsection{Fitness criteria}
Most optimization algorithms including GAs work best when the entire fitness of a member can be encapsulated into one parameter or expression. Multi-parameter optimization techniques are usually difficult to implement and slow to converge. However, the choice of the fitness parameter must be such that a gradual increase in fitness will lead to a closer approach to an optimum solution \cite{ga1}. Moreover, the fitness landscape (the pictorial representation of various candidate solutions as hills and vallies) must have one clear global optimum and not too many local optima. GAs tend to get trapped in local optima \cite{ga1,ga3}, and a fitness function with many vallies of the same depth may lead to the GA getting trapped in one of them without finding the deepest valley (global optimum).\\
We used three kinds of fitness functions depending on the target matrix $U$. Let $G$ be the candidate solution (one member of the population), and $U$ be the target
\begin{eqnarray}
F_1 &=& \frac{1}{\rm{Tr}|G^{\dagger}U - U^{\dagger}U|}\\
F_2 &=& \frac{1}{\sum_{i=1}^{64}|G^{\dagger}U - U^{\dagger}U|_{i}}\\
F_3 &=& \frac{1}{\sum_{i=1}^{64}|G - U|_{i}}\\
\end{eqnarray}
where the sum in $F_2$ and $F_3$ indicates the sum over all 64 elements of the matrix $|G^{\dagger}U - U^{\dagger}U|$ and $|G - U|$. Function $F_1$ works best when the target $U$ is sparse, and this is usually true for all quantum gates of interest. However, when $U$ is diagonal and many elements are $1$, it tends to fall into a local optimum trap of $G=\bf{1}_{8\times 8}$ ie the unity operator. For example in implementing the $\Lambda_2(I_z)$ gate [2,18,19]
\[\Lambda_2(I_z) = \left(\begin{matrix} 1 & 0 & 0 & 0 & 0 & 0 & 0 & 0\\ 0 & 1 & 0 & 0 & 0 & 0 & 0 & 0 \\ 0 & 0 & 1 & 0 & 0 & 0 & 0 & 0 \\0 & 0 & 0 & 1 & 0 & 0 & 0 & 0 \\0 & 0 & 0 & 0 & 1 & 0 & 0 & 0 \\0 & 0 & 0 & 0 & 0 & 1 & 0 & 0 \\0 & 0 & 0 & 0 & 0 & 0 & 1 & 0 \\ 0 & 0 & 0 & 0 & 0 & 0 & 0 & -1 \end{matrix}\right)\]
the solution employing $F_1$ tends to get trapped into $\bf{1}_{8\times 8}$.\\
In such cases $F_2$ is found to work better. $F_3$ works best when the matrix $U$ is not sparse. This has limited application in most useful quantum gates, but it can be used in running a local-optimizer.

\subsection {Details of convergence and algorithm parameters}
We worked with population sizes ranging from 500 to 1000. Larger population sizes lead to better convergence to the solution but take longer time. A single run of 50 generations takes about 5-7 minutes on a 2.6GHz AMD personal computer using MATLAB. The relatively short times involved, and considering that we use only a standard PC make this method attractive. The run process is as follows: given the matrix $U$, we fix the number of rows of the population members as 3 to start with. The entries of the population are initially taken to be random numbers from 0-9. Then we run over 50 generations, and if $U$ can be decomposed using only 3 steps, the solution usually appears as the member having highest fitness at the end of the run. To explore if there are any other decompositions of $U$ using only 4 steps, the process can be repeated. However, there is no guarantee the solution may appear, as the algorithm may tend to converge to one of the solutions more predominantly.\\
The number of rows is increased to 4, and the whole process is repeated, again to 50 generations. By repeating this and sweeping the number of rows (this corresponds to the number of steps $N$ we seek to decompose $U$ into), from 3-10, we obtain various possible decompositions of $U$.\\
However, it is not ensured that \textit{all} decompositions of $U$ appear as a result of this algorithm. But most target matrices of interest cannot be decomposed in too many different ways, and the algorithm still yields useful and more efficient pulse sequences than those currently being employed.

\subsection{Local Optimizer}
If an exact solution is found, then the fitness according to (15) should be infinite. However, since we are doing calculations in MATLAB using fixed point arthimetic, it is usually about $10^{15}$. Since the accuracy of the $\theta_i$ in the above GA is only to $5^{\circ}$, the best solutions usually have fitness of about 1000.  To approach the exact solution from here, we use a local optimizer. This can be done in two ways:
\begin{enumerate}
\item Use a "greedy hill climbing" algorithm that sweeps each $\theta_i$ in the range $\pm 10^{\circ}$, and finds the best solution.\\
\item Use another genetic algorithm where to all the $\theta_i$ are added random numbers in the range $\pm 10^{\circ}$. By using a large population size (2000), and using fitness function $F_3$, and only employing cross-over operation CROSS-4 but allowing the fourth column entries to be floating point numbers, we find a solution to within about $0.1^{\circ}$ accuracy in 5 minutes. \\
\end{enumerate}

\section{Results}
In this section, we present some results of our algorithm in finding efficient sequences for the inversion on equality gate, parity gate and fanout gate, and compare them with traditional sequences.
\subsection{Inversion on Equality Gate}
Inversion-on-equality ($I_=$) gate inverts the state when the states of all the qubits are equal. In a
3-qubit system it act as \cite{anil1}
\begin{equation}
I_{=}|abc> = (-1)^{\delta_{ab} \delta_{bc}}|abc>
\end{equation}
The unitary operator for implementation of this gate is of the form
\begin{equation}
U_{I_=} = \left(\begin{matrix} -1 & 0 & 0 & 0 & 0 & 0 & 0 & 0\\ 0 & 1 & 0 & 0 & 0 & 0 & 0 & 0 \\ 0 & 0 & 1 & 0 & 0 & 0 & 0 & 0 \\0 & 0 & 0 & 1 & 0 & 0 & 0 & 0 \\0 & 0 & 0 & 0 & 0 & 1 & 0 & 0 \\0 & 0 & 0 & 0 & 1 & 0 & 0 & 0 \\0 & 0 & 0 & 0 & 0 & 0 & 1 & 0 \\ 0 & 0 & 0 & 0 & 0 & 0 & 0 & -1 \end{matrix}\right)
\end{equation}
\begin{figure}
\includegraphics[width=12cm]{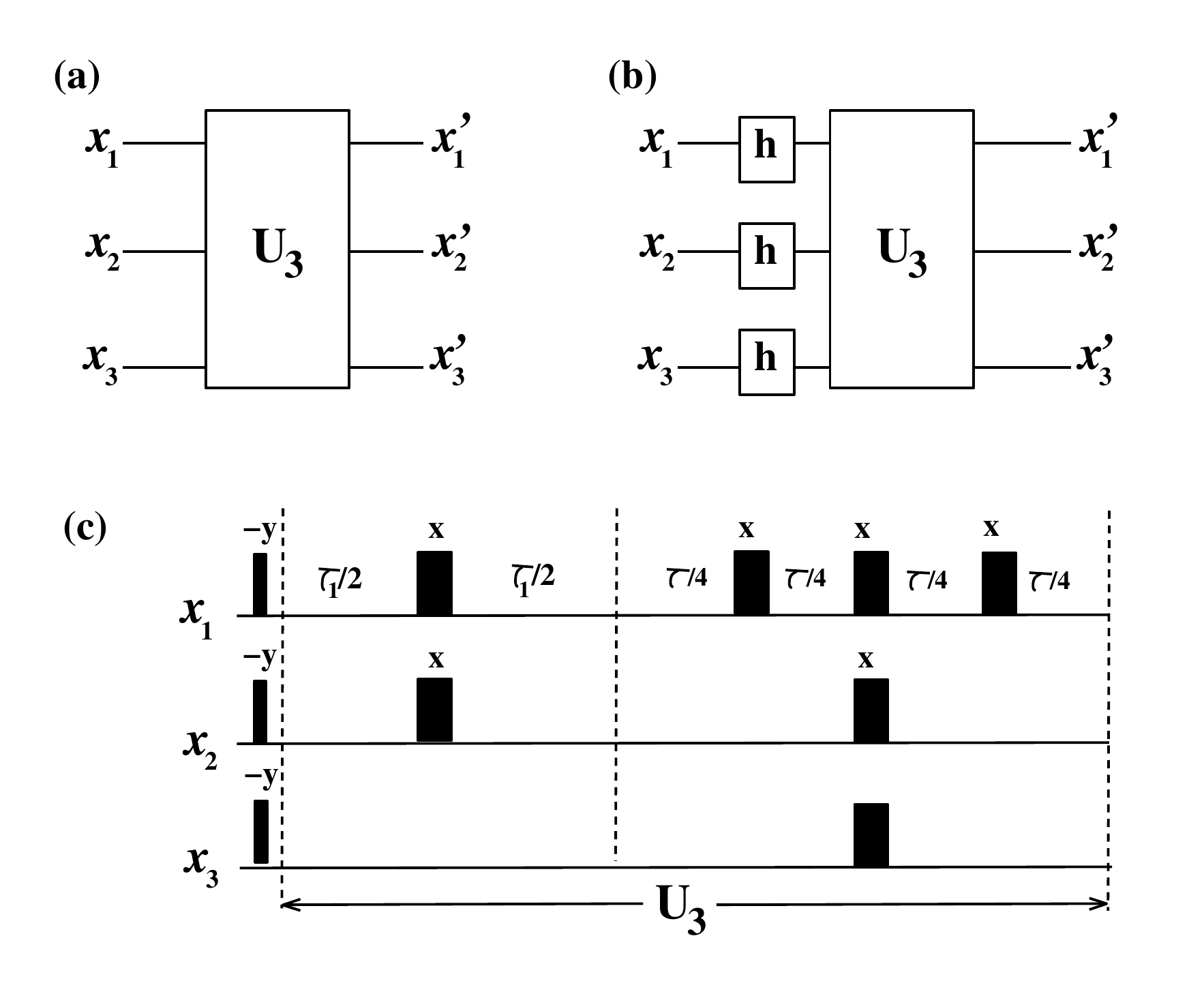}
\caption{The traditional implementation of the inversion on equality gate when $J_{12}, J_{23}$ and $J_{13}$ exist. Here $\tau_1=1/2J_{12}$, $\tau_2=1/2J_{23}$ and $\tau_3=1/2J_{13}$.}
\end{figure}
Applying $I_=$ on a three qubit system, when one of the qubits is in state $|1>$, results in controlled-Z gate in the other two qubits. Such gates are used in the implementation of universal CNOT gates.
The well known sequence for this is \cite{anil1,anil2}
\begin{equation}
U_{I_{=}} = \exp\left(-i2\pi J_{12}\tau_1 I_{1z}I_{2z}\right) \exp\left(-i2\pi J_{23}\tau_2I_{2z}I_{3z}\right) \exp\left(-i2\pi J_{31}\tau_3I_{1z}I_{3z}\right) 
\end{equation}
where $\tau_1=1/2J_{12}$, $\tau_2=1/2J_{23}$ and $\tau_3=1/2J_{13}$.
Our program yields this same sequence as the best possible. However, this is in the case when all three coupling constants $J_{12}, J_{23}$ and $J_{13}$ exist. What will be the optimal sequence in oft-considered case when $J_{13}=0$ and $J_{12}=J_{23}=J$?\\


The advantage of automating the process is that it just requires a small change in the code. We replace all the bilinear propagators $\exp(-i\theta I_{1\alpha}I_{3\beta})$ with the the combined propagator of form $\exp\left(-i\theta (I_{1z}I_{2z} + I_{2z}I_{3z}) \right)$. This encapsulates the fact that $J_{13}=0$, as the bilinear propagators involving $I_{1}$ and $I_3$ will not be implementable. Our program leads in this case the sequence
\begin{eqnarray}
U_{I_{=}} &=& \exp\left(-i\frac{\pi}{4}\bf{1}\right)\exp\left(-i2\pi I_{1z}I_{2x}I_{3y}\right)\exp\left(-i\pi(I_{1z}I_{2z} + I_{2z}I_{3z})\right)\nonumber\\
&\times&\exp\left(-i2\pi I_{1z}I_{2y}I_{3x}\right)\exp\left(-i\pi I_{2y}I_{3y}\right)
\end{eqnarray}
This requires a total period of $\tau=2.73/J$. We believe this is the best way of realizing $U_{I_=}$ in this case.
\subsection {Parity Gate}
Parity gate adds (addition modulo 2) the control bits to the target bit \cite{xiv1,xiv2}. The parity gate on a 3-qubit system where third qubit is the target qubit and the first two qubits are control, has the form

\begin{equation}
P = \left(\begin{matrix} 1 & 0 & 0 & 0 & 0 & 0 & 0 & 0\\ 0 & 1 & 0 & 0 & 0 & 0 & 0 & 0 \\ 0 & 0 & 0 & 1 & 0 & 0 & 0 & 0 \\0 & 0 & 1 & 0 & 0 & 0 & 0 & 0 \\0 & 0 & 0 & 0 & 0 & 1 & 0 & 0 \\0 & 0 & 0 & 0 & 1 & 0 & 0 & 0 \\0 & 0 & 0 & 0 & 0 & 0 & 1 & 0 \\ 0 & 0 & 0 & 0 & 0 & 0 & 0 & 1 \end{matrix}\right)
\end{equation}

\begin{figure}
\includegraphics[width=16cm]{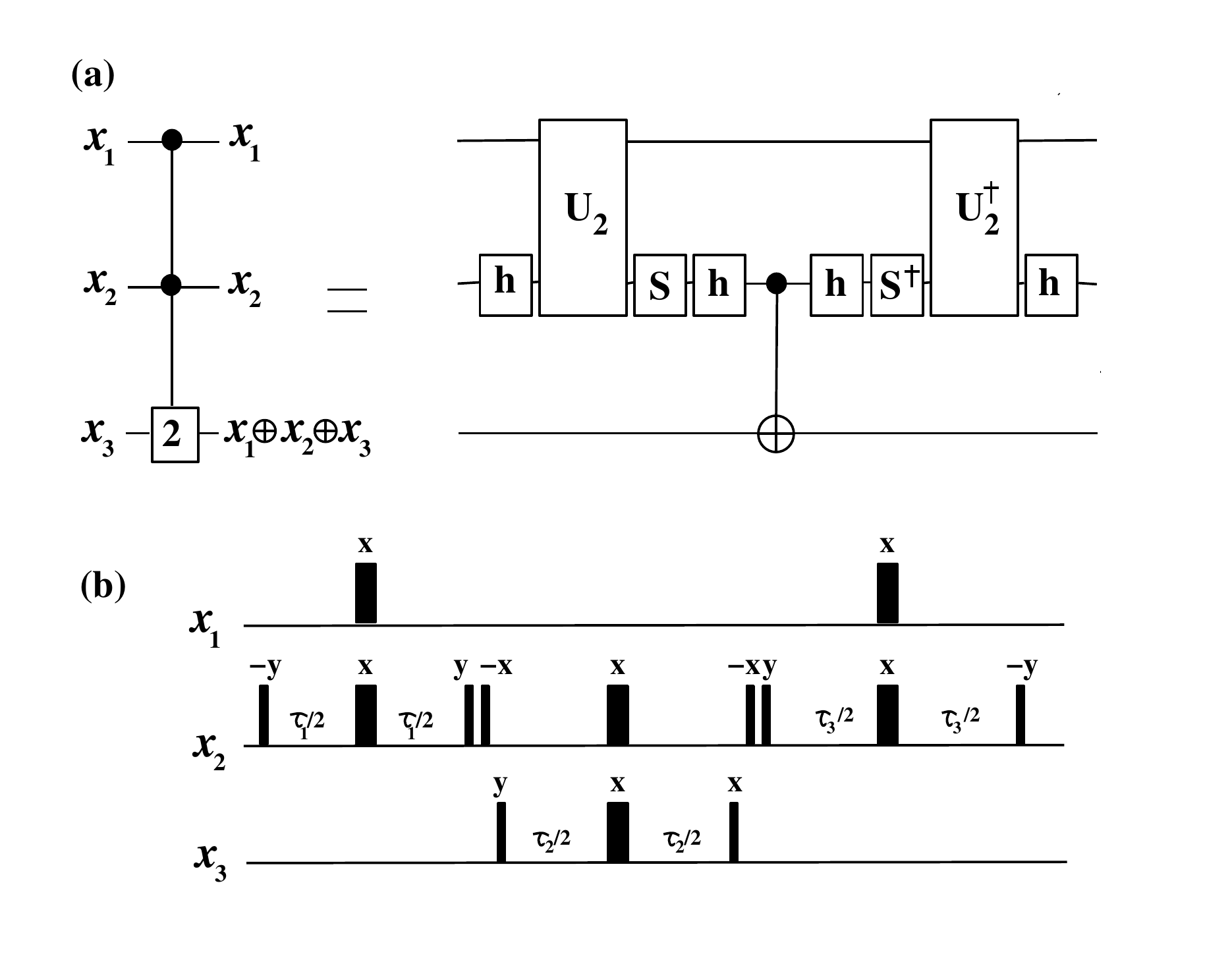}
\caption{The traditional implementation of the parity gate requiring a total time period of $\tau=2.5/J$. Here $\tau_1=1/2J_{12}$, $\tau_2=1/2J_{23}$ and $\tau_3=1/2J_{13}$.}
\end{figure}

The conventional implementation uses a two-qubit $U_2$ gate which is of the form \cite{anil1,anil2} $U_2=e^{-i2\pi(JI_{1z}I_{2z})}\tau$ where $\tau=1/2J$.$U^{\dagger}$ is of the same form but with $\tau=3/2J$. Besides $U_2$ and pseudo-Hadamard gates, the other gates used are phase gate and CNOT gate given by
\begin{equation}
s = \left(\begin{matrix} 1 & 0 \\ 0 & i \end{matrix}\right)
\end{equation}
\begin{equation}
CNOT = \left(\begin{matrix} 1 & 0 & 0 & 0 \\ 0 & 1 & 0 & 0 & \\ 0 & 0 & 0 & 1 \\0 & 0 & 1 & 0 \end{matrix}\right)
\end{equation}

The sequence and circuit are shown in fig 4. The total time required is $\tau = 2.5/J$.\\
Using our improved genetic algorithm, we obtain the following decomposition for the parity gate

\begin{equation}
P =\exp\left(-i\frac{\pi}{4}\bf{1}\right) \exp\left(-i\frac{3\pi}{2}I_{3x}\right)\exp\left(i\pi I_{1z}I_{2z}\right)\exp\left(-i2\pi I_{1z}I_{2z}I_{3x}\right)
\end{equation}
requiring a total period of $1.366/J$, which is 54\% faster than the conventional sequence \cite{anil1}. Moreover, this sequence is found to require only about 25\% of the RF power of the conventional sequence.

\subsection {Fanout Gate}

\begin{figure}
\includegraphics[width=16cm]{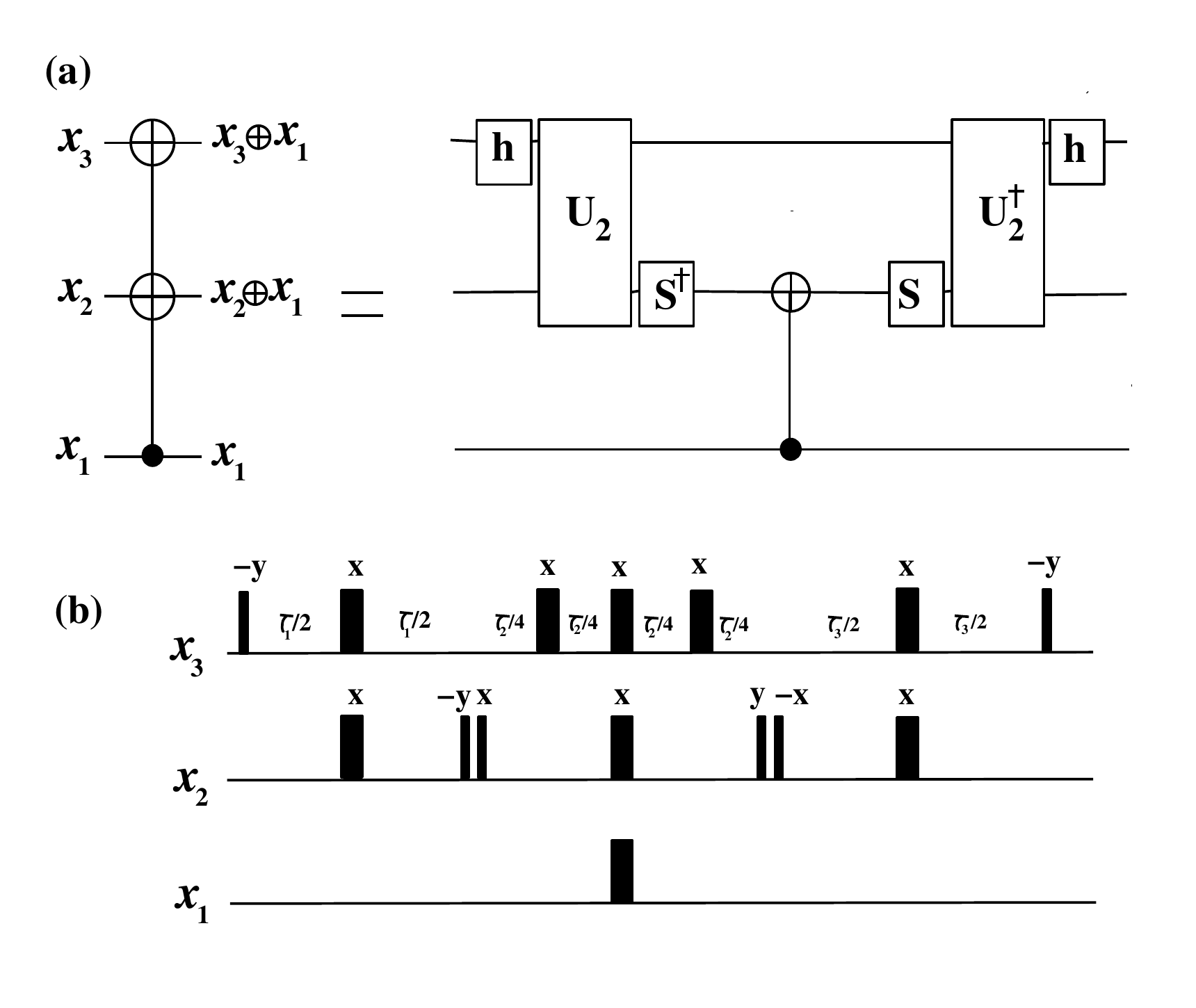}
\caption{The traditional implementation of the fanout gate requiring a total time period of $\tau=2.5/J$. Here $\tau_1=1/2J_{12}$, $\tau_2=1/2J_{23}$ and $\tau_3=1/2J_{13}$.}
\end{figure}

Fanout gate adds (addition modulo 2) a control bit onto 'n' target bits \cite{xiv1,xiv2,anil22}.The fanout gate on a 3-qubit system where first qubit is the control qubit and other two qubits are target is shown below. The (classical) value of control bit is copied or 'fanned out' to the target bits if the target bits are initially in $|0>$ state. However, if the control bit is in coherent superposition, the fanout gate creates entangled states. A (n+1) fanout gate is conventionally built out of (n+1) parity gate by applying Hadamard gates on both sides of the parity gate on the first n-qubits \cite{anil1,anil2}
\begin{equation}
F = \left(\begin{matrix} 1 & 0 & 0 & 0 & 0 & 0 & 0 & 0\\ 0 & 1 & 0 & 0 & 0 & 0 & 0 & 0 \\ 0 & 0 & 1 & 0 & 0 & 0 & 0 & 0 \\0 & 0 & 0 & 1 & 0 & 0 & 0 & 0 \\0 & 0 & 0 & 0 & 0 & 0 & 0 & 1 \\0 & 0 & 0 & 0 & 0 & 0 & 1 & 0 \\0 & 0 & 0 & 0 & 0 & 1 & 0 & 0 \\ 0 & 0 & 0 & 0 & 1 & 0 & 0 & 0 \end{matrix}\right)
\end{equation}

The conventional sequence requires a time period of $\tau=2.5/J$. Our algorithm yields
\begin{equation}
F = \exp\left(-i\frac{\pi}{4}\bf{1}\right) \exp\left(i\pi I_{2x}I_{3x}\right)\exp\left(-i2\pi I_{1z}I_{2x}I_{3x}\right)\exp\left(-i\frac{3\pi}{2}I_{1z}\right)
\end{equation}

requiring $\tau=1.366/J$, which is 54\% faster than the conventional period.

\section{Conclusion}
In this paper, we have developed a new algorithmic way of decomposing a target matrix $U$ into hard pulses, bilinear and trilinear propagators. By automating this process and finding various possible decompositions of a matrix, we show that one can determine more time efficient pulse sequences to realize the same gate $U$ for a 3-spin NMR problem. This is then applied to determining efficient pulse sequences for the invert-on-equality gate $U_{I_=}$, the parity gate $P$ and fanout gate $F$. In the last two cases the sequences are about 50\% faster than conventional sequences, and require lesser RF power for their implementation. \\
Although this is a marginal improvement, it is interesting that the same algorithm is able to yield both sequences by only changing the target matrix $U$. We foresee the use of such an algorithmic technique when either the pulse sequence for a gate is unknown (as in $U_{I_=}$, or if it is too difficult to intuitively determine.

\section{appendixes}
We present here the flow-charts describing the traditional genetic algorithms, and our improved genetic algorithm.

\begin{figure}
\includegraphics[width=16cm]{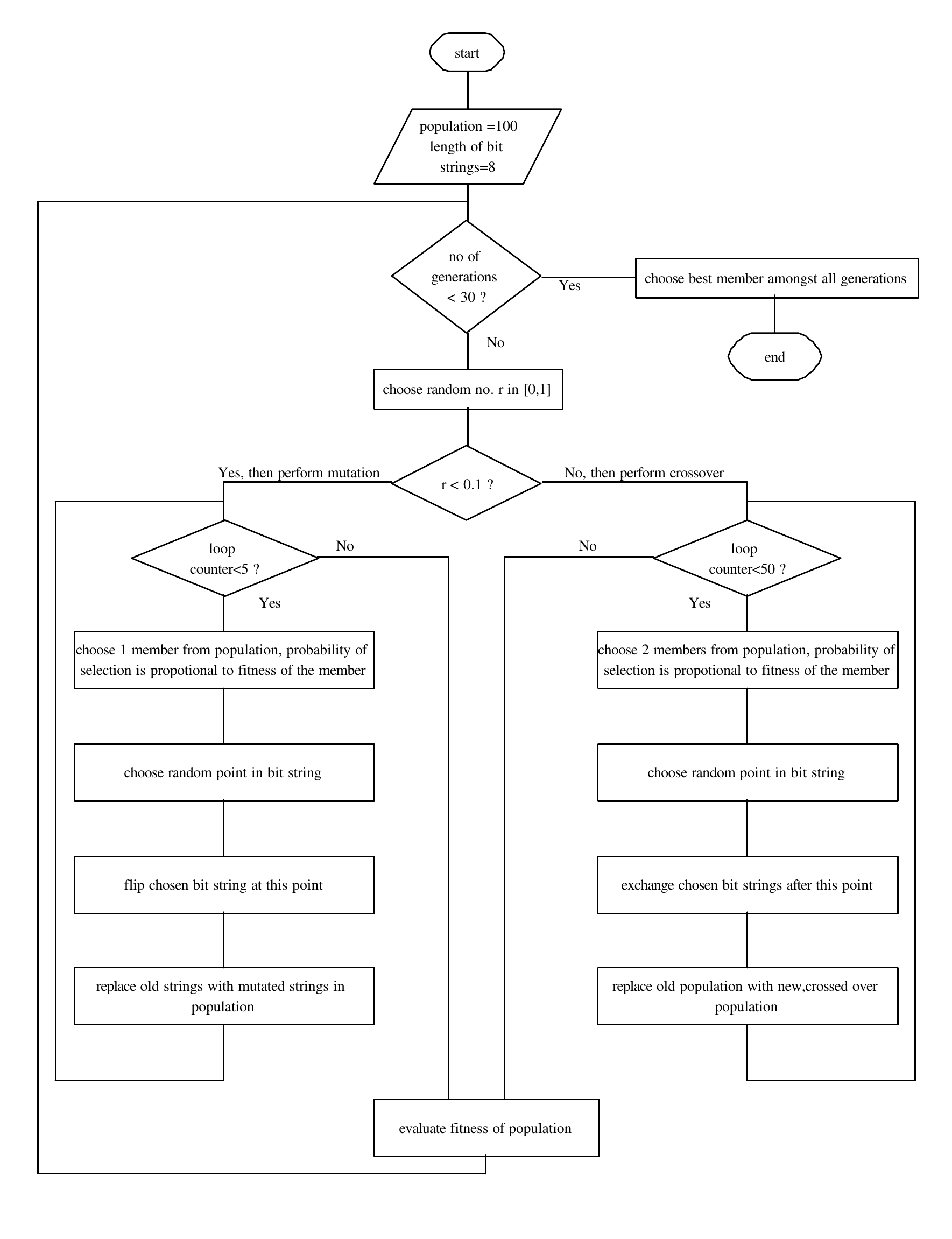}
\caption{The traditional genetic algorithm employing bit-strings and single crossover and mutation operations. For definiteness, we fix the population size as 100, bit-string length as 8 bits and the number of reproduction cycles (generations) as 30 in this flowchart.}
\end{figure}

\begin{figure}
\includegraphics[width=14cm]{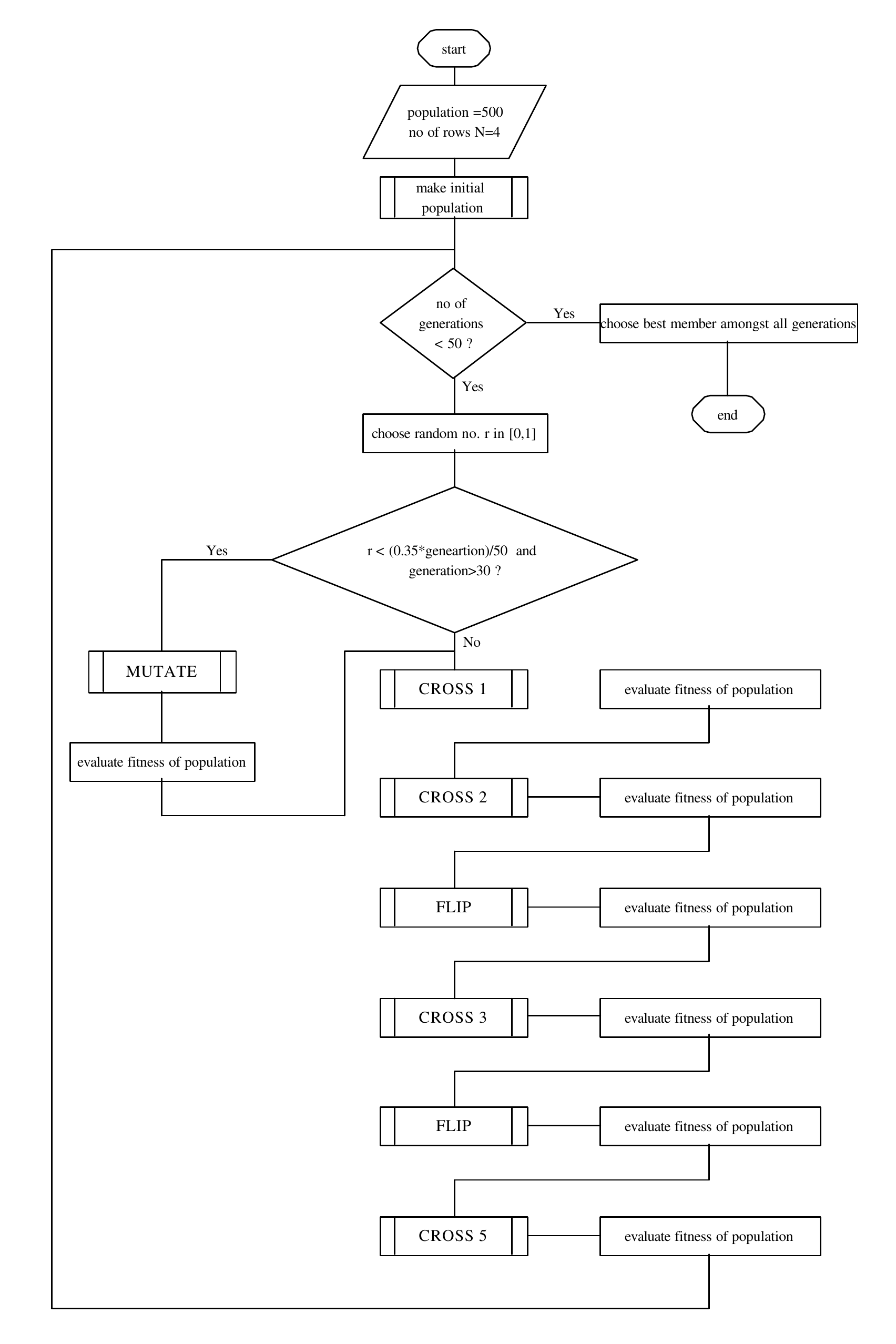}
\caption{The improved genetic algorithm employing matrices and a mutation and 5 types of crossover operations. We typically used population sizes as 500, and number of generations as 50. This flowchart illustrates the case when $N=4$. The program is swept from $N=4$ to $N=10$.}
\end{figure}

\begin{figure}
\includegraphics[width=16cm]{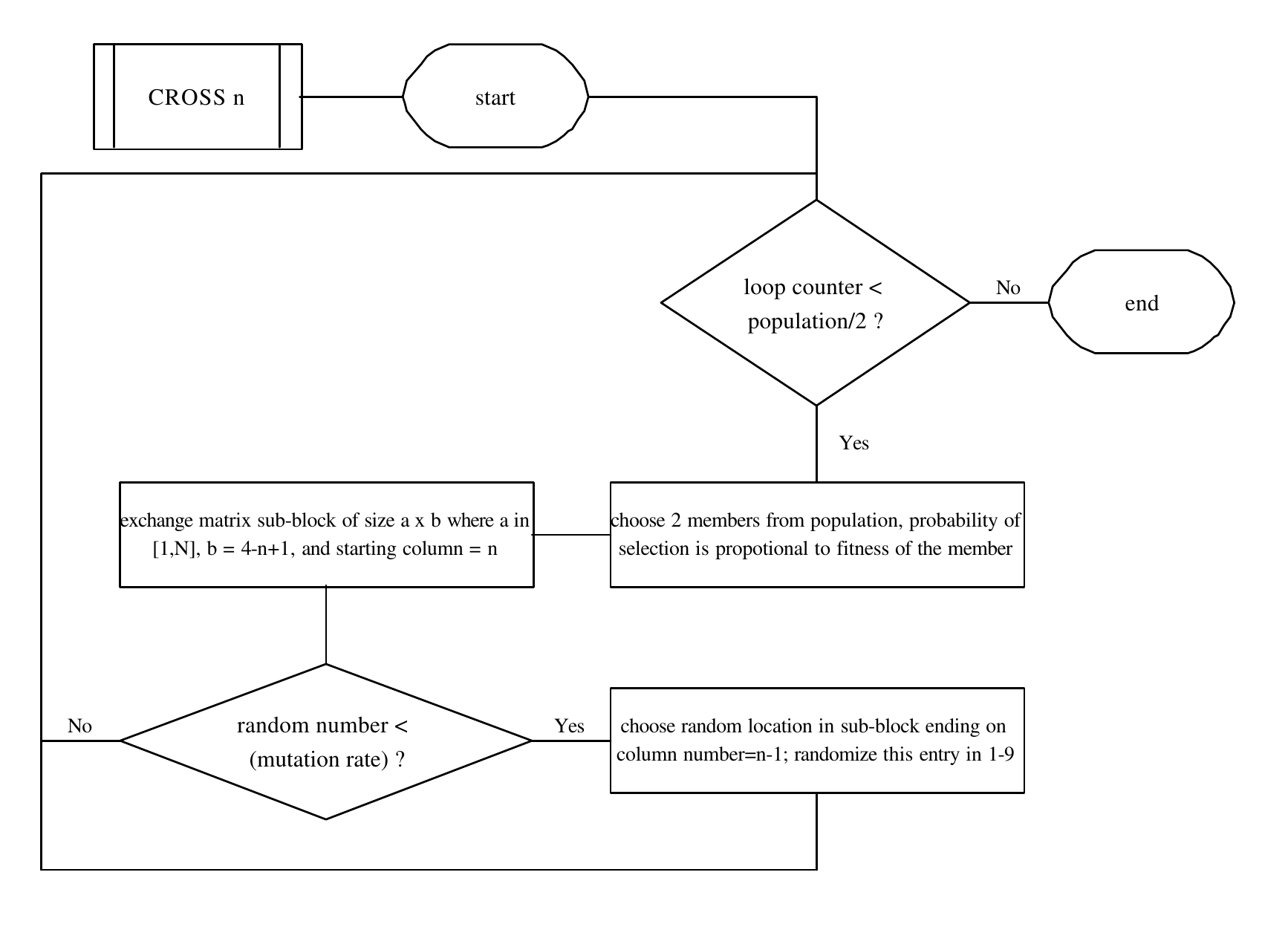}
\caption{The generic crossover operation CROSS n where $n\in [1,4]$.}
\end{figure}

\begin{figure}
\includegraphics[width=16cm]{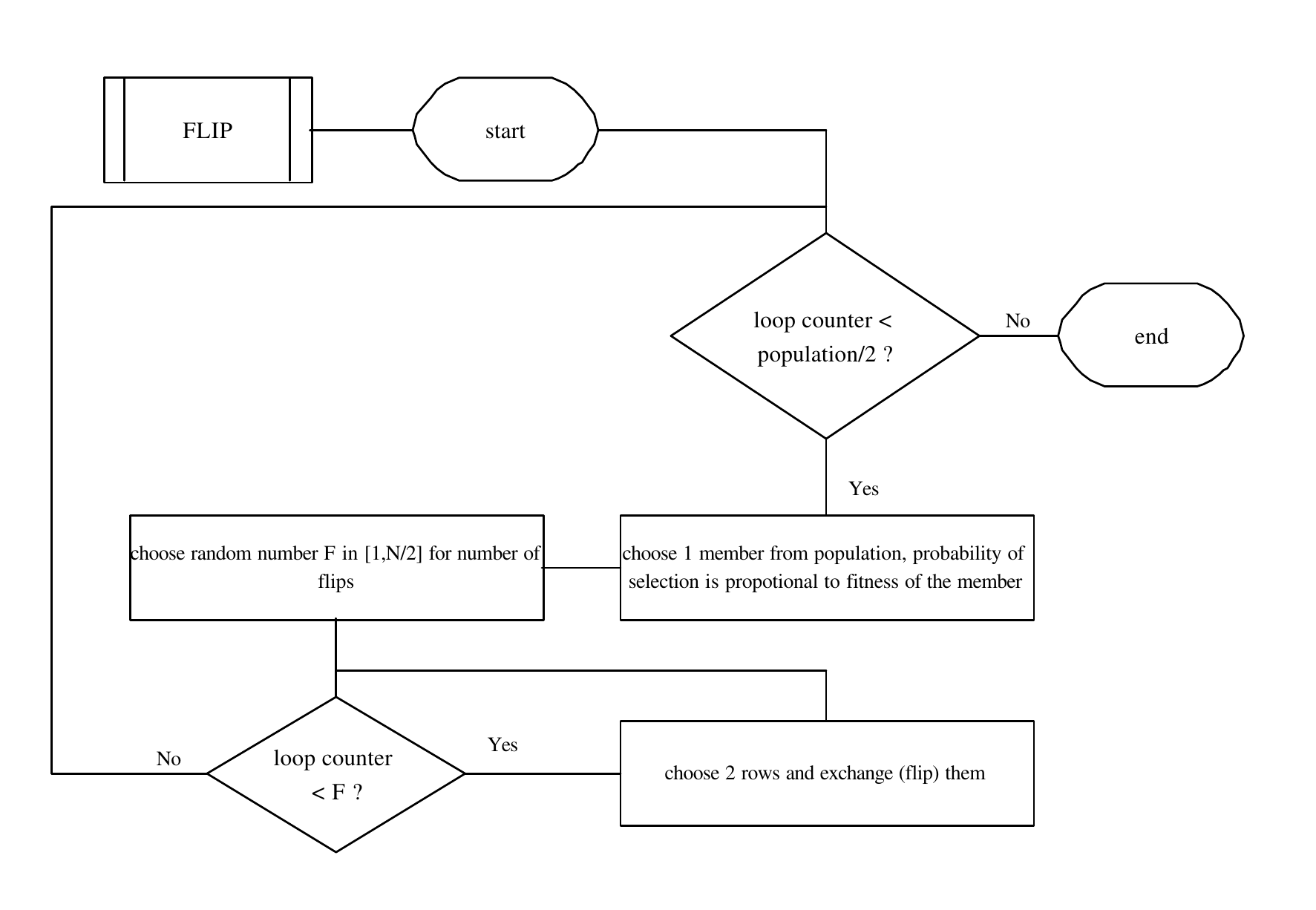}
\caption{The FLIP operation. This is introduced to account for non-commutativity of suceeding rows (propagators) in the member matrices.}
\end{figure}

\end{document}